\documentclass{frontiersFPHY} 

\usepackage{url,hyperref,lineno,microtype}
\usepackage[onehalfspacing]{setspace}

\def\keyFont{\fontsize{8}{11}\helveticabold }
\def\firstAuthorLast{Bryce Yoshimura {et~al.}} 
\def\Authors{Bryce Yoshimura\,$^{1,*}$ and J. K. Freericks\,$^{1}$ }
\def\Address{$^1$Department of Physics, Georgetown University, Washington , DC, USA  }
\def\corrAuthor{Bryce Yoshimura}
\def\corrAddress{Department of Physics, Georgetown University, 37th and O Sts. NW, Washington, DC, 20057, USA}
\def\corrEmail{bty4@georgetown.edu}

\begin{document}
\onecolumn
\firstpage{1}

\title[Determining the ground-state probability $\ldots$]{Determining the ground-state probability of a quantum simulation with product-state measurements} 

\author[\firstAuthorLast ]{\Authors} 
\address{} 
\correspondance{} 

\extraAuth{}

\maketitle


\begin{abstract}

\section{}
One of the goals in quantum simulation is to adiabatically generate the ground state of a complicated Hamiltonian by starting with the ground state of a simple Hamiltonian and slowly evolving the system to the complicated one. If the evolution is adiabatic and the initial and final ground states are connected due to having the same symmetry, then the simulation will be successful. But in most experiments, adiabatic simulation is not possible because it would take too long, and the system has some level of diabatic excitation. In this work, we quantify the extent of the diabatic excitation even if we do not know {\it a priori} what the complicated ground state is. Since many quantum simulator platforms, like trapped ions, can measure the probabilities to be in a product state, we describe techniques that can employ these measurements to estimate the probability of being in the ground state of the system after the diabatic evolution. These techniques do not require one to know any properties about the Hamiltonian itself, nor to calculate its eigenstate properties. All the information is derived by analyzing the product-state measurements as functions of time.

\tiny
 \keyFont{ \section{Keywords:} Quantum simulation, ion trap, adiabatic state preparation, transverse field Ising model, ground state probability } 
\end{abstract}

\section{Introduction}

%
%

The Hilbert space that describes a strongly correlated many-body quantum system grows exponentially in the number of particles $N$, so determining the ground state of a complex many-body quantum system becomes numerically intractable when the size of the quantum system becomes too large to be represented on a classical computer (unless there is some other simplification, like weak entanglement, {\it etc.}). The idea to simulate complex many-body quantum systems on a quantum computer was proposed by Feynman in the 1980's~\cite{feynman1981}. Since Feynman's proposal quantum alogorithms have been proposed to calculate eigenvalues and eigenvectors of these intractable systems~\cite{lloyd1999}.One of the challenges with creating the ground state on a quantum computer, say by adiabatic evolution of the ground state from a simple to a complex one, is how do we determine the extent of the ground-state preparation. After all, we don't know what the ground state is {\it a priori} so it is difficult to know what the probability to be in the ground state is. In this work, we propose one method to determine the probability to remain in the ground state. While this analysis is applied to  ion-trap emulators (that model interacting spin systems), the general discussion can be applied to any quantum computer that performs ground state preparation, but creates diabatic excitations as a result of a too rapid time evolution. To date, trapped-ion quantum simulators have seen success in two different platforms: the Penning trap has trapped $\approx 300$ ions in a planar geometry and generated Ising spin interactions~\cite{britton2012} and the linear Paul trap has performed quantum simulations with $18$ ions in a one-dimensional crystal~\cite{senko2013}. The success of these traps as quantum simulators is attributed to their long coherence times, precise spin control, and high fidelity. Here, we will focus on the linear Paul trap quantum simulator.    

In an ion trap quantum simulator, hyperfine states of the trapped ions are used for the different
spin states (for simplicity, we can consider only two states, and hence, a spin-one-half system). Optical pumping can be employed to create a product state with the ions all in one of the two hyperfine states with fidelities close to 100\%.  A coherent rotation of that state can then be used to create a wide range of different initial product states. By turning on a large magnetic field, this state can configured to be the ground state of the system. Then the magnetic field is reduced slowly enough for the system to remain in the ground state until the system evolves into the complex spin Hamiltonian in zero magnetic field. The challenge is that the evolution of the system must be completed within the coherence time of the spins, which often is too short to be able to maintain adiabaticity throughout the evolution (and indeed, becomes increasingly difficult as the system size gets larger). We propose to employ the time evolution of an observable, $\mathcal{O}(t)=\langle\psi(t)|\hat{\mathcal{O}}|\psi(t)\rangle$ (with $|\psi(t)\rangle$ the quantum state of the system at time $t>t_{stop}$), after the evolution to the final Hamiltonian, to measure the absolute probability of the ground state. The time evolution of the observable, $\mathcal{O}(t)$, oscillates at frequencies given by the energy differences between the final eigenstates (where the Hamiltonian becomes time independent). More concretely, let $\hat{\mathcal{H}}(t_{stop}) |m\rangle = E_m|m\rangle$,   then the time-dependent expectation value satisfies
\begin{equation}
	\mathcal{O}(t>t_{stop}) = \sum_{mn} P_m^* P_n \langle m|\hat{\mathcal{O}} |n\rangle \exp[-i(E_n - E_m)t],
\label{eq:oscillations}
\end{equation}
where $P_m = \langle m| \psi(t_{stop}) \rangle $ is the overlap of the state $| \psi(t_{stop}) \rangle$ with the eigenstate $|m\rangle$ (we have set $\hbar=1$). Previously, we showed how Fourier transforming the time series and employing
signal processing methods like compressive sensing, allows one to extract the energy differences as a type of many-body eigenstate spectroscopy~\cite{shabani2011, spectro_us}. Here, we focus on the amplitude of the oscillations,
given by $P^*_1P_n$, which is proportional to the probability amplitude of the ground state ($P_1$), if the ground state still has a high probability in $|\psi(t_{stop})\rangle$. Note that we do not need to know the explicit ground-state wavefunction to extract its probability from these oscillations. This is the main advantage of this technique.

\begin{figure}[h!]
	\begin{center}
		\includegraphics[scale=0.08]{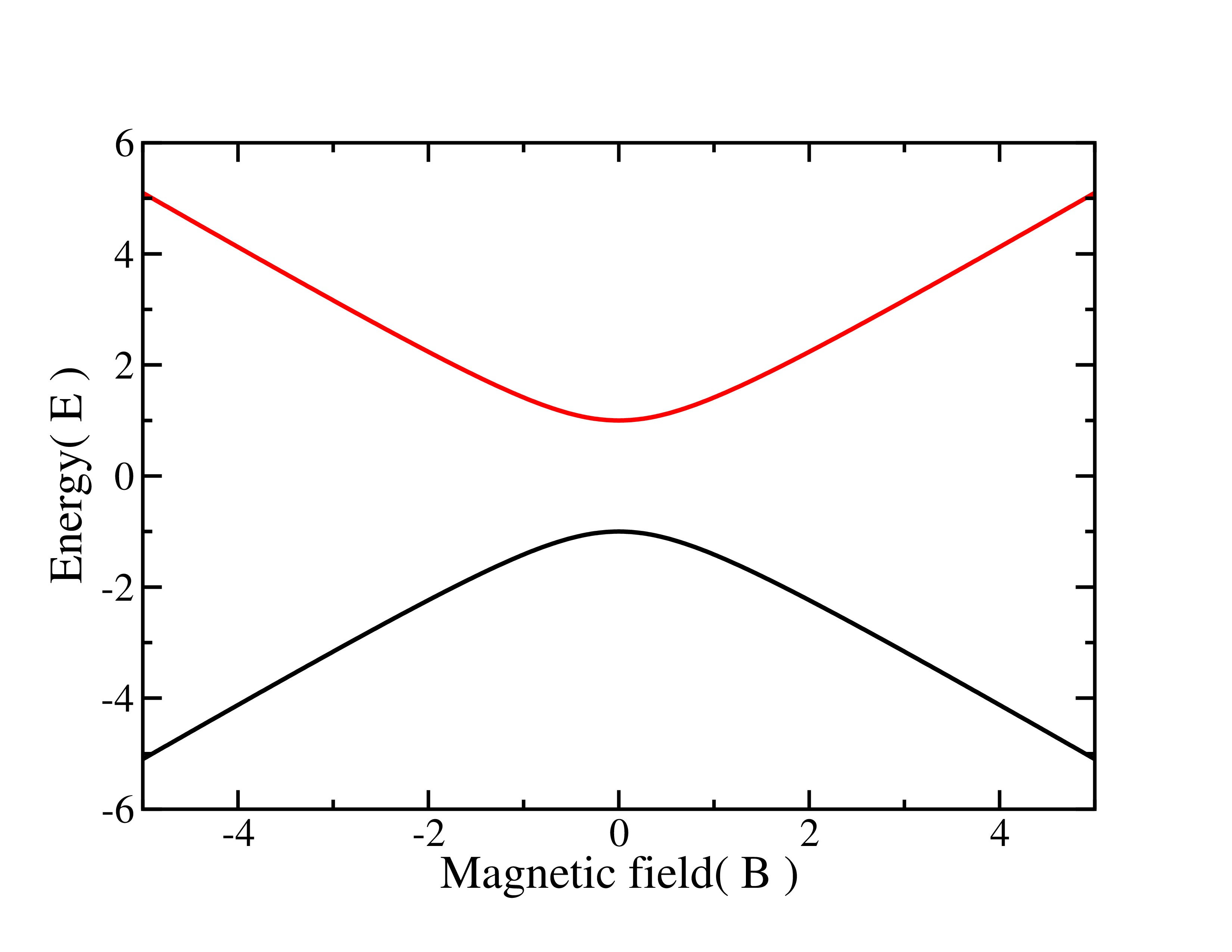}
	\end{center}
	\textbf{\refstepcounter{figure} \label{fig:lzenergy} Figure \arabic{figure}.} {(Color online.) The energy spectra of the Landau-Zener problem. There is a minimum energy gap when the excited state, in red, lies at its minimum energy difference with the ground state. This occurs at $B^{z}(t) = 0$. }	
\end{figure}

\begin{figure}[h!]
	\begin{center}
		\includegraphics[scale=0.08]{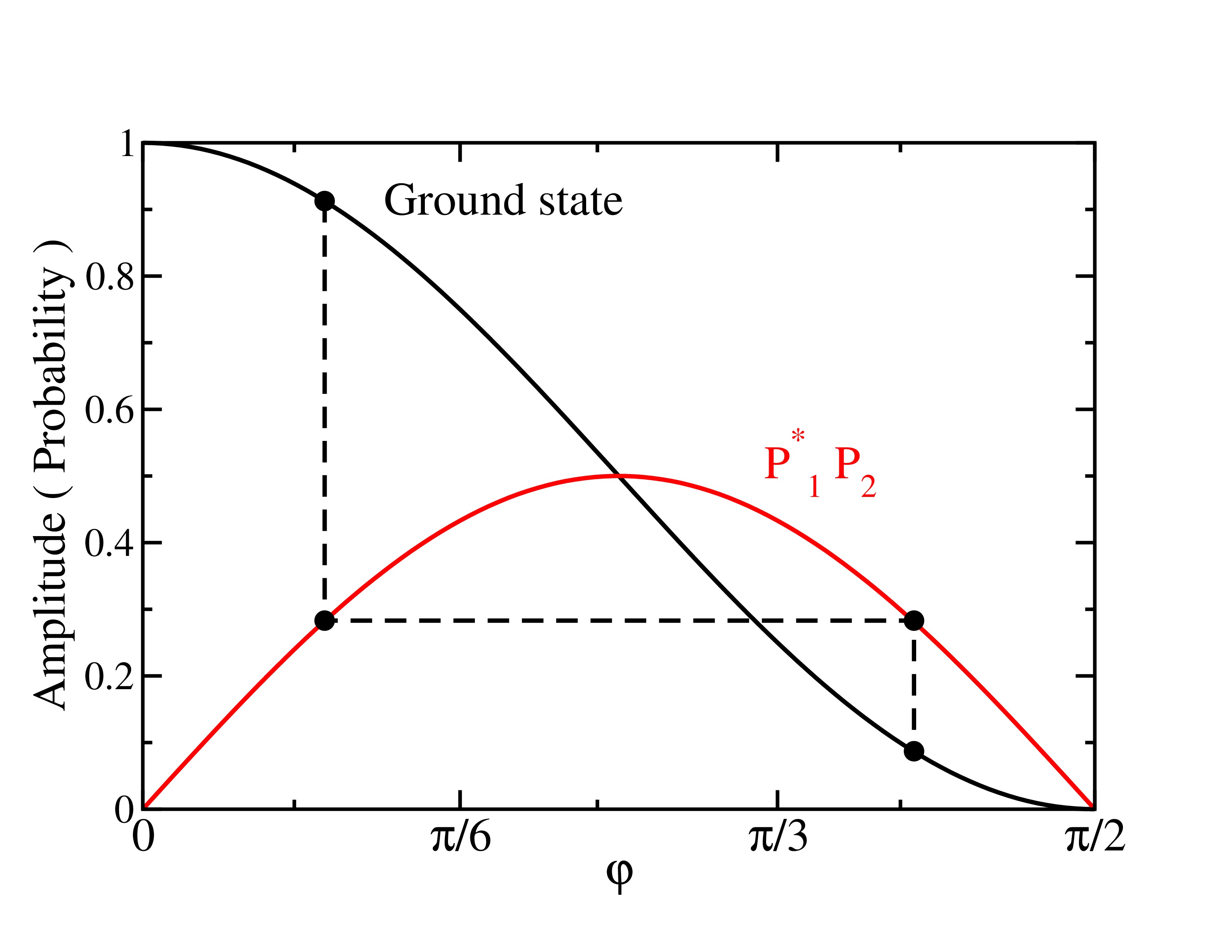}
	\end{center}
	\textbf{\refstepcounter{figure} \label{fig:exact} Figure \arabic{figure}.} {(Color online.) Analytic solutions to the Landau-Zener problem given ground-state probability (black solid line) and the amplitude of the oscillations, $P^*_1P_2$, of $\hat{\mathcal{O}}(t > t_{stop})$ (red solid line). The amplitude of the oscillations is not unique to a single ground-state probability, as highlighted by the dashed line. } 
\end{figure}

We illustrate how the the ground state probability can be extracted  by analyzing the amplitude of the oscillations of  the simplest time-dependent Hamiltonian: the two-level Landau-Zener problem.  The Landau-Zener problem is defined via
\begin{equation}
	\hat{\mathcal{H}}(t) = B^{z}(t) \sigma^{z} +  \sigma^{x}.
\end{equation}
Here $\sigma^{\alpha}$ are the Pauli spin operators in the $\alpha = x$, $y$, or $z$ direction. The Pauli spin operators have the commutation relation
\begin{equation}
	\left[\sigma^{\alpha}, \sigma^{\beta} \right] = 2i \epsilon_{\alpha \beta \gamma} \sigma^{\gamma}, 
	\label{eq:spincommutation}
\end{equation}
where the Greek letters represent the spatial directions and $\epsilon_{\alpha \beta \gamma}$ is the antisymmetric tensor. The Landau-Zener problem has a minimum energy gap occurring when $B^{z}(t) = 0$, as shown in Fig.~\ref{fig:lzenergy}.
Since the Landau-Zener problem is a two state system, the probabilities, $P_1$ and $P_2$, are related by $P_2^2 = 1 - P_1^2$ and the state, $|\psi(t_t) \rangle$, can be represented by $P_1 = \cos(\phi)$ and $P_2 = \sin(\phi)$. Using this state in Eq.~(\ref{eq:oscillations}), we find that the expecation value  $\mathcal{O}(t > t_{stop})$ becomes (neglecting terms with no time dependence)
\begin{equation}
	\mathcal{O}(t>t_{stop}) = \cos(\phi)\sin(\phi) \left \{ \langle 1|\hat{\mathcal{O}} |2\rangle \exp[-i(E_2 - E_1)t] + \langle 2|\hat{\mathcal{O}} |1\rangle \exp[-i(E_1 - E_2)t] \right \}.
\label{eq:LZoscillations}
\end{equation}
The ground-state probability [$\cos^2(\phi)$] can then be calculated if $\langle 1|\hat{\mathcal{O}} |2\rangle$ is known. The amplitude of the oscillations is $\sin( 2\phi )\langle 1|\mathcal{O}|2\rangle$. However, even if $\phi$ is extracted from the amplitude, there are two always solutions, except when $\phi = \pi/4$ (see Fig.~\ref{fig:exact}), and hence two possible ground state probabilities. In Fig.~\ref{fig:exact} we show both the amplitude of the oscillations and the ground-state probability as a function of $\phi$, where the dashed line shows that the amplitude is not unique to a single ground-state probability. However, once the system has more states, one does not have a simple closed set of equations and the analysis of the amplitude of the oscillations can only deduce the ground-state probability when the ground-state amplitude is dominant in $|\psi(t)\rangle$. We demonstate this below with the transverse field Ising model.   

\begin{figure}[h!]
	\begin{center}
	\begin{tabular}{c}
		\includegraphics[scale=0.27]{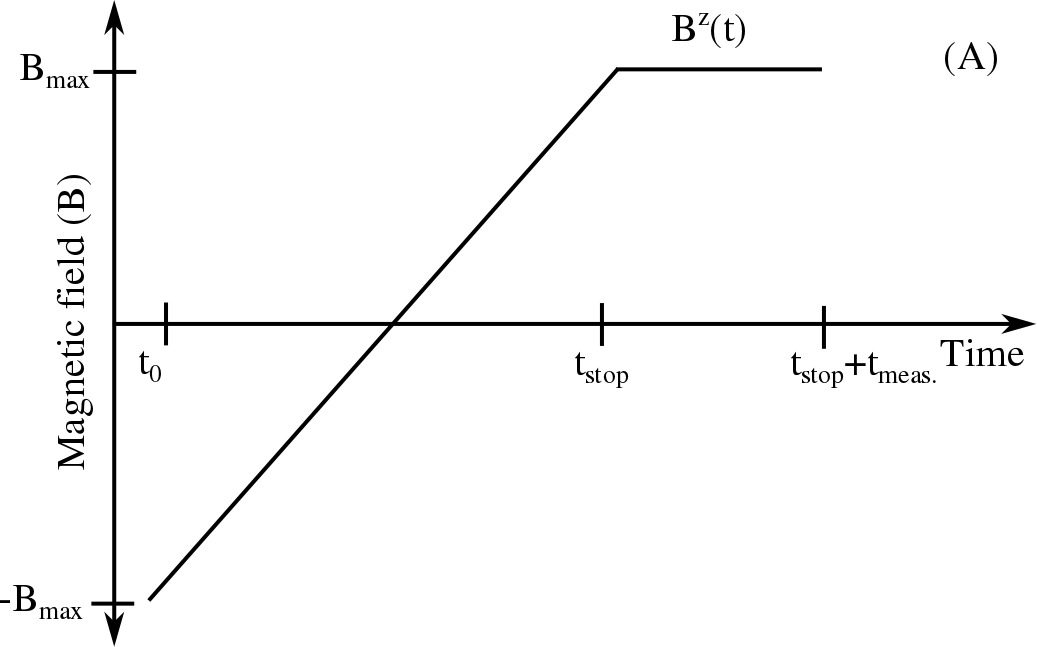}\\
		\hspace{10pt}
		\includegraphics[scale=0.087]{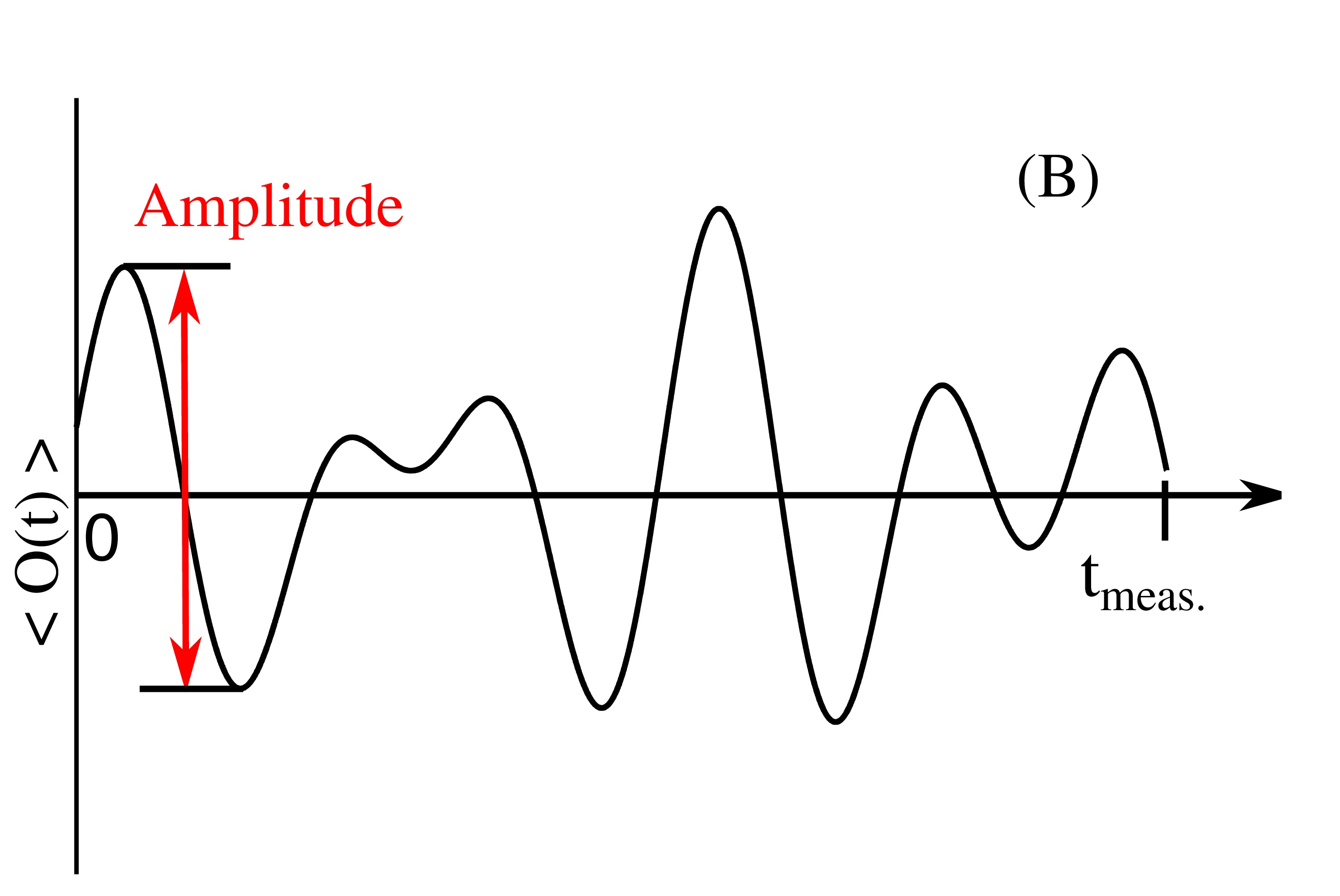}
		\end{tabular}
	\end{center}
	\textbf{\refstepcounter{figure} 	\label{fig:lzschematic} Figure \arabic{figure}.} {(Color online.) Schematic diagram of the experimental protocol. (A) The transverse magnetic field as a function of time is diabatically ramped down to a chosen value, $B^{z}(t_{stop})$. $B^{z}(t_{stop})$ is then held for a time interval of $t_{meas.}$. (B) An observable is measured at the end of the interval $t_{meas.}$. The
amplitude of the oscillation is taken, for simplicity, to be the amplitude of the initial oscillation after $t_{stop}$. }
\end{figure}

It is well known that the amount of diabatic excitation in the Landau-Zener problem increases the faster 
the magnetic field is ramped from $-\infty$ to $+\infty$. The general protocol that we employ is as follows (and is depicted schematically in Fig.~\ref{fig:lzschematic}):     
\begin{enumerate}
	\item Initialize the state in an arbitrary state, in the following examples we initialize the state in the ground state of the Hamiltonian with a large polarizing magnetic field.
	\item Decrease the magnetic field as a function of time to evolve
	 the quantum state, as shown in Fig.~\ref{fig:lzschematic}(A), where, for concreteness,  we show an example of a magnetic field that changes linearly.
	\item Hold the magnetic field at its final value which is first reached at $t = t_{stop}$ until the measurement is performed at  the time interval $t_{meas.}$ after the field has been held constant [see  Fig.~\ref{fig:lzschematic}(A)].
	\item Measure an observable of interest, $\mathcal{O}(t)$, for a number of different $t_{meas.}$ values.
	\item Determine the amplitude of the oscillations.
\end{enumerate}

Note that one requirement of this approach is that
the observable of interest must oscillate as a function in time as given in Eq.~(\ref{eq:oscillations}). The amplitude is extracted from the first maximum and minimum of the observable as a function of time by 
\begin{equation}
	Amplitude = \frac{ \text{max}[\mathcal{O}(t)] - \text{min}[\mathcal{O}(t)]}{2}
\end{equation}

The time evolution of the wavefunction $|\psi(t)\rangle$ is calculated by solving the time-dependent Schr\"odinger equation:
\begin{equation}
	i\frac{\partial}{\partial t} | \psi(t) \rangle = \hat{\mathcal{H}}(t) | \psi(t) \rangle
\end{equation}
by using the Crank-Nicolson method to time evolve the state $| \psi (t)\rangle$. This technique solves the problem with the following approach~\cite{cranknicolson}
\begin{equation}
	\left( \hat{ \mathbb{I}} + i \frac{\delta t}{2} \hat{ \mathcal{ H } }( t + \delta t) \right) | \psi (t + \delta t) \rangle = \left( \hat{\mathbb{I}} - i \frac{\delta t}{2} \hat{ \mathcal{ H } }( t ) \right) | \psi (t ) \rangle. 
\end{equation}
Note that the Hamiltonian is time-dependent until $t_{stop}$ is reached, when it becomes constant in time.

We present a numerical example to illustrate this protocol by analyzing the oscillations for the Landau-Zener problem. Due to the fact that the eigenstates for the Landau-Zener problem at $|B^{z}(t_{stop})| \gg 1$ approach the eigenstates of the $\sigma^z$ operator, if one measures an operator that is diagonal in this basis, there will be no oscillations in the expectation value.  Hence, we measure the expectation value of the operator $\hat{\mathcal{O}}(\theta)$, the Pauli spin matrix that points in the $\theta$ direction. 
\begin{equation}
	\hat{\mathcal{O}}(\theta) =  R^{\dagger}(\theta)  \sigma^{z} R(\theta),  
\end{equation} 
where $R(\theta)$ is the global rotation about the $y$-axis and is given by
\begin{equation}
	R(\theta) = \hat{\mathbb{I}} \cos\left(\frac{\theta}{2} \right) + i \sigma^{y} \sin\left(\frac{\theta}{2} \right), 
\end{equation}
where $\theta = \pi/2$ produces $\hat{\mathcal{O}}(\theta=\pi/2)= \sigma^{x}$. 

\begin{figure}[h!]
	\begin{center}
	\includegraphics[scale=0.08]{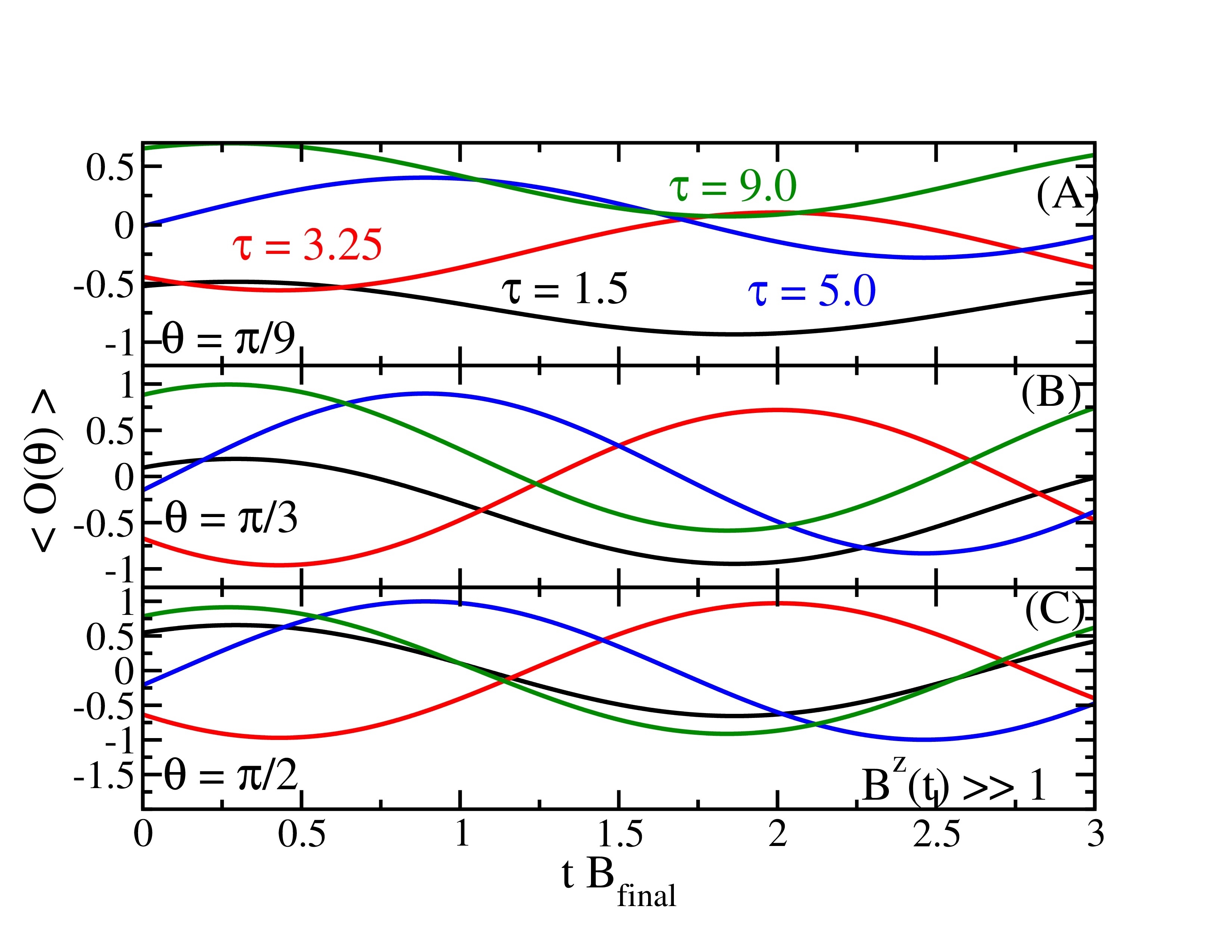}
	\end{center}
	\textbf{\refstepcounter{figure} \label{fig:lzsignal} Figure \arabic{figure}.} {(Color online.) Time evolution of the measurement of $\hat{\mathcal{O}}(\theta)$ for $\tau= 1.5$ (black), $3.25$ (red), $5.0$ (blue), and $9.0$ (green) using $\theta = \pi/6$ (A), $\pi/3$  (B), and $\pi/2$ (C). The measurement is being done when $B^{z}(t) \gg 1$. The amplitude of the oscillations converge to $1$ as $\theta$ is increased to $\pi/2$. }
\end{figure}
For our numerical examples with the Landau-Zener problem, we use a linear ramp, $B^{z}(t) = \tau t + B_0$, where $B_0 < 0$. $|B_0|$ is chosen to be large in comparison to 1 to polarize the spin. We evolve the state to $t_{stop}$, such that $B^{z}(t_{stop}) \gg 1$. We present the time evolution for $4$ different $\tau = 1.5$, $3.25$, $5.0$, and $9.0$ for $3$ different $\theta = \pi/9$, $\pi/3$, and $\pi/2$ in Fig.~\ref{fig:lzsignal}. The amplitude of the oscillations becomes $1$ when $\theta = \pi/2$. When $\tau=5.0$ the amplitude of the oscillations is maximized in comparison to the $3$ other $\tau$'s.          

\begin{figure}[h!]
	\begin{center}
	\includegraphics[scale=0.08]{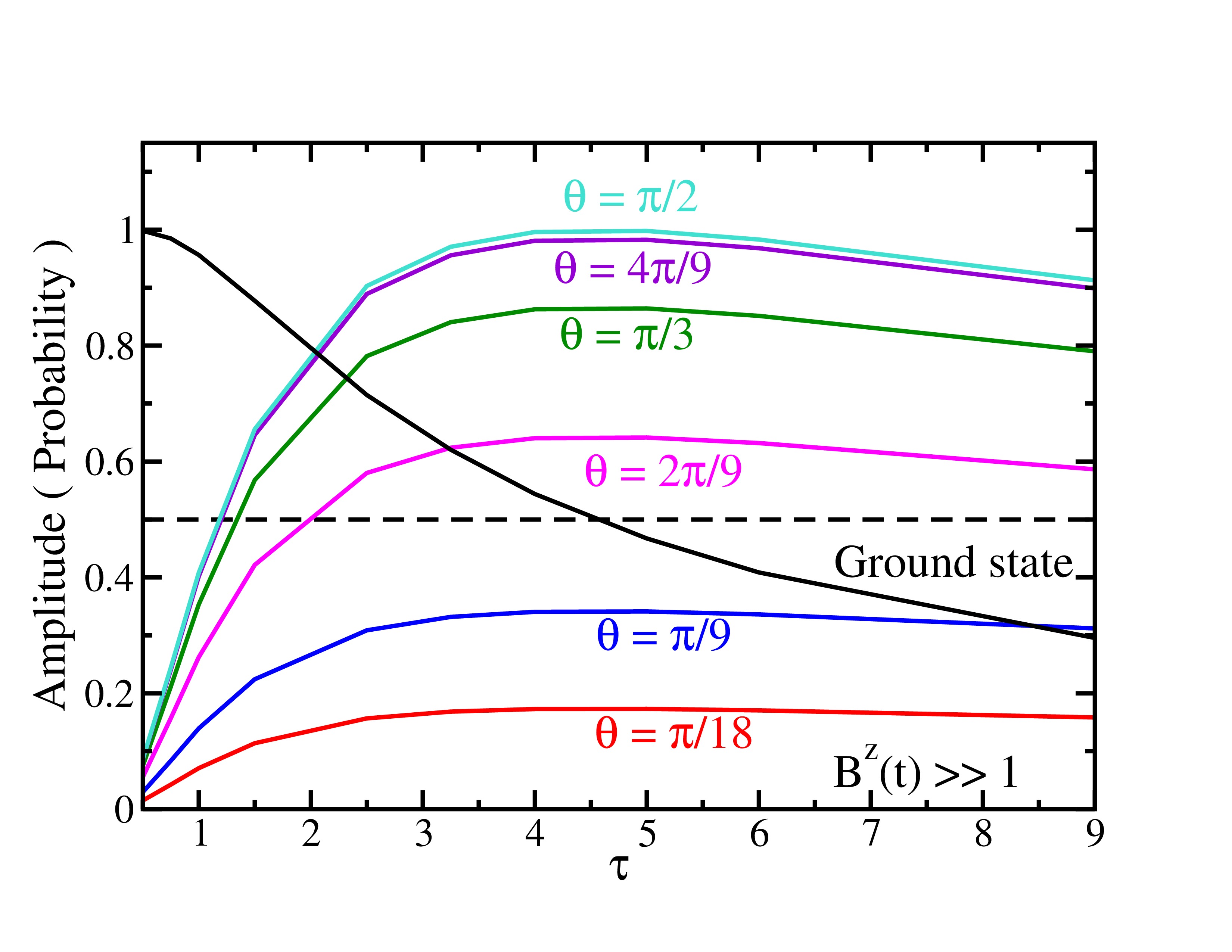}
	\end{center}
	\textbf{\refstepcounter{figure} \label{fig:lzamplitude} Figure \arabic{figure}.} {(Color online.) Analysis of the amplitude of the oscillation as a function of the $\tau$ for $6$ different $\theta$'s. The dashed line is when the ground-state probability is $0.5$. As $\theta$ increases to $\pi/2$ the amplitude scale increases as well. When the ground-state probability is near $0.5$ the amplitude of the oscillations is maximized and the amplitude decreases when either the ground state probability decreases or increases. }
\end{figure}

In Fig.~\ref{fig:lzamplitude}, we show the probability of the ground state compared to the amplitude of the oscillations as a function of $\tau$. The amplitude of the oscillations increases as $\theta$ is increased. This is due to the $\sigma^{x}$ term dominating the $\hat{\mathcal{O}}(\theta)$ operator instead of the $\sigma^{z}$ term. When the ground state probability approaches $0.5$ the amplitude of the oscillations is maximal and it decreases either when the ground state probability increases or decreases. As the probability of ground state approaches $1$ the amplitude is expected to become $0$, which can be obscured due to experimental noise. In order to determine which side of the maximum the measurement of the amplitude is on, multiple $\tau$ experiments must be run to track the depletion of the ground state as the amplitude of the oscillations reaches a maximum and then decreases.  Note further that in this case, since there is only one excited state, one can, in principle, always determine the ground-state probability by
measuring the amplitude of the oscillation and extracting the appropriate probability amplitude. For more complex systems, such a procedure will not be possible, but the monotonic nature of the
curve (at least while the probability for the ground state remains above 50\%) will allow us to determine whether a given run of the experiment increases the probability to be in the ground state, which can be employed to optimize the ground-state preparation if it is done with some alternative quantum control method besides adiabatically evolving the system. Indeed, we believe this has the potential to be the most important application of this approach.

This simple example shows us a number of important points. First, one may need to rotate the measurement basis if the final product basis are eigenvectors of the Hamiltonian. Second, as the ground state is depleted, the amplitude of the oscillations grows until it reaches a maximum, when the system is equally populating both eigenstates. If the ground state is further depleted, the amplitude of the oscillations will decrease. One can make a mistake in estimating the probability in the ground state if one does not know which side of the curve one is on (probability of the ground state below or above 50\%). On the other hand, if one knows which side of the curve one is on, due to making measurements at earlier times to track the ground-state depletion, then one might be able to further use the amplitude to determine the ground state probability in the Landau-Zener problem.  When we change to the ion-trap system and examine the transverse-field Ising model, then the procedure becomes more complicated because there are more states that the ground state can be depleted into, and this complicates the analysis.


\section{Transverse field Ising model} 

\begin{figure}[h!]
	\begin{center}
	\includegraphics[scale=0.08]{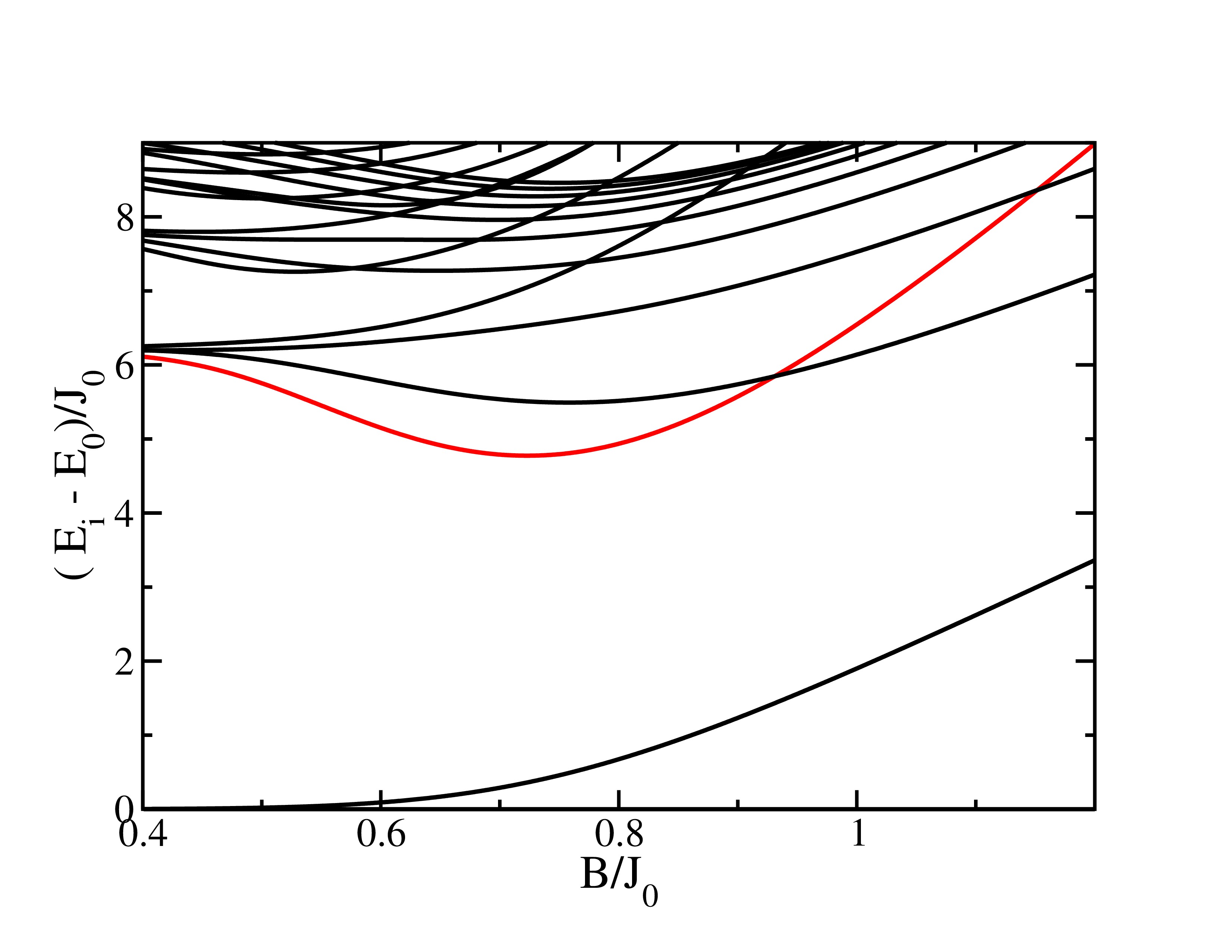}
	\end{center}
	\textbf{\refstepcounter{figure} \label{fig:isingenergy} Figure \arabic{figure}.} {(Color online.) The energy spectra of the ferromagnetic transverse field Ising model with N=10. The red curve shows the first coupled excited state and the minimal gap occurs at $B^x\approx 0.72$. }
\end{figure}

Now we describe a more realistic case of the transverse field Ising model. The transverse field Ising model for $N$ particles is given by 
\begin{equation}
	\hat{\mathcal{H}}(t) = -J_{\pm} \sum^N_{i < j} J_{ij}\sigma^{z}_i \sigma^{z}_j - B^{x}(t)\sum^N_{i=1} \sigma^{x}_i,
	\label{eq:ComHam}
\end{equation}
where the $J_{ij}$ are the spin-spin interactions produced by a spin-dependent force and given by~\cite{monroe_duan}
\begin{equation}
	J_{ij} = \Omega^2\nu_R \sum_{\nu =1}^N \frac{b^*_{i\nu} b_{j\nu}}{\mu^2 - \omega^2_{\nu}},
	\label{eq:interaction}
\end{equation}
where $b_{i\nu}$ is the normalized eigenvector of the $\nu^{\rm th}$ phonon modes, $\omega_{\nu}$ is the corresponding frequency of the phonon mode, $\Omega$ is the single spin flip Rabi frequency, and $\nu_R$ is the recoil frequency associated with the dipole force on an ion, from which we define our energy units with $J_0 = \Omega^2\nu_R$ . The Raman beatnote frequency $\mu$ is tuned to the blue of the largest $\omega_{\nu}$ (which here is the center-of-mass phonon, $\omega_{COM}$). The details of calculating the $J_{ij}$ can be found elsewhere ~\cite{three_ion}. The $J_{ij} \propto | r_{ij}|^{-\alpha} $ with $r_{ij}$ the interparticle distance and the exponent $\alpha$ being tunable between 0 and 3. The exponent $\alpha$ is tuned by changing $\mu$ or by changing the ratio of the longitudinal to the transverse trap frequencies. Here, we study the ferromagnetic interaction of the Ising model with $J_{\pm} > 0$. The Pauli spin matrices are now associated with each lattice site. The $J_{ij}$ of the transverse Ising model have spatial-reflection symmetry such that $J_{ij} = J_{ji}$ and the eigenstates have the same symmetry. The eigenstates of the transverse field Ising model also have spin-reflection parity, that is, under the partial inversion transformation $\sigma^{x} \rightarrow \sigma^{x}$, $\sigma^{y} \rightarrow -\sigma^{y}$, and $\sigma^{z} \rightarrow -\sigma^{z}$. The spin-reflection parity and spatial-reflection symmetry produce avoided crossings between eigenstates with the same parity and symmetry, such that a minimum energy gap to the lowest coupled state occurs as shown in Fig.~\ref{fig:isingenergy}.

\begin{figure}[h!]
	\begin{center}
		\begin{tabular}{ c }
			\includegraphics[scale=0.27]{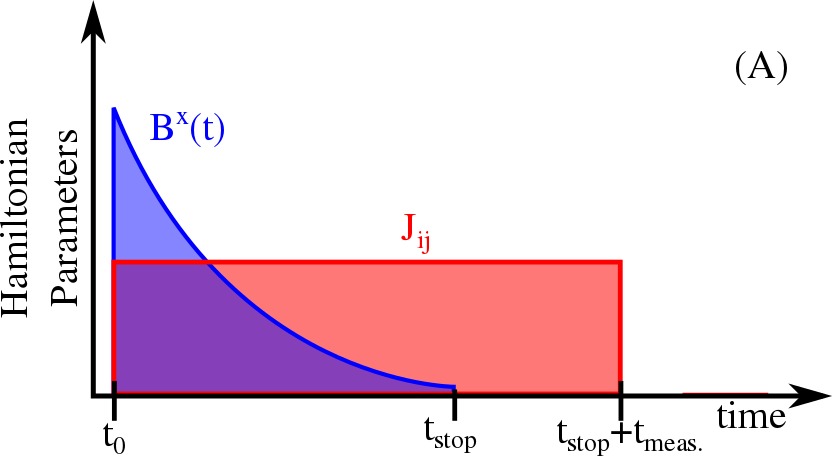}\\
			\includegraphics[scale=0.27]{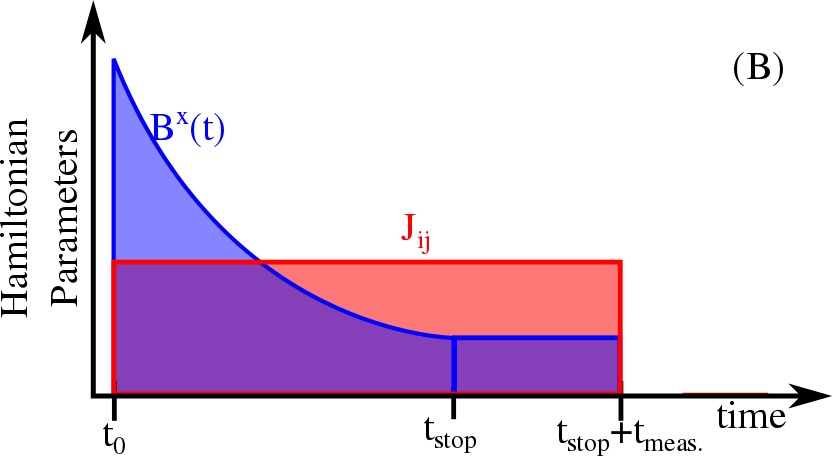}
		\end{tabular}
	\end{center}
	\textbf{\refstepcounter{figure} \label{fig:Isingcartoon} Figure \arabic{figure}.} {(Color online.) Schematic diagram of the experimental protocol used for the transverse field Ising model. (A) The transverse magnetic field as a function of time is diabatically ramped down to a chosen value, $B^{x}(t_{stop} = 6\tau)$. Then $B^x(t)$ is quenched to $0$ before the start of the measurement time interval, $t_{meas.}$. (B) The transverse magnetic field is ramped down and held constant at a final value determined by $t=t_{stop}$. }
\end{figure}
The experimental protocol is essentially the same as before, except for a few differences. The first difference being that the transverse magnetic field now depends exponentially as a function of time and is given by
\begin{equation}
	B^{x}(t) = B_o \exp^{-\frac{t}{\tau}},
\end{equation}
as shown in Fig.~\ref{fig:Isingcartoon}(A). The second difference is that we will perform two different experimental protocols where the initial state is evolved to $t_{stop} = 6\tau$ and before the time interval $t_{meas.}$ starts, the magnetic field is quenched to zero $B^{x}(t_{meas.})=0$, as shown in Fig.~\ref{fig:Isingcartoon}(A), or the transverse magnetic field is held at its final value which is first reached at $t = t_{stop}$, as depicted in Fig.~\ref{fig:Isingcartoon}(B). We work with parameters for the ion chain where the exponent $\alpha\approx 1$. The energy spectra are plotted in Fig.~\ref{fig:isingenergy}.

There are a number of additional complications. First off, the eigenstates at $B^x=0$ are product states along the $z$ direction, hence we need to rotate again to see the oscillations. We choose $\hat{\mathcal{O}}(\theta)$ to be the average magnetization in the $\theta$-direction
\begin{equation}
	\hat{\mathcal{O}}(\theta) = \frac{1}{N}  R^{\dagger}(\theta) \sum_{i=1}^N \sigma_i^{z} R(\theta),  
\end{equation} 
where $R(\theta)$ is now the global rotation given by
\begin{equation}
	R(\theta) = \prod_{i=1}^N \left [ \hat{\mathbb{I}} \cos\left(\frac{\theta}{2} \right) + i \sigma^{y}_i \sin\left(\frac{\theta}{2} \right) \right ] , 
\end{equation}
where $\theta = \pi/2$ yields $\sigma^{x}_{\rm tot}$. Measuring the average magnetization in the $\theta$-direction produces the needed oscillations.

\begin{figure}[h!]
	\begin{center}
	\includegraphics[scale=0.08]{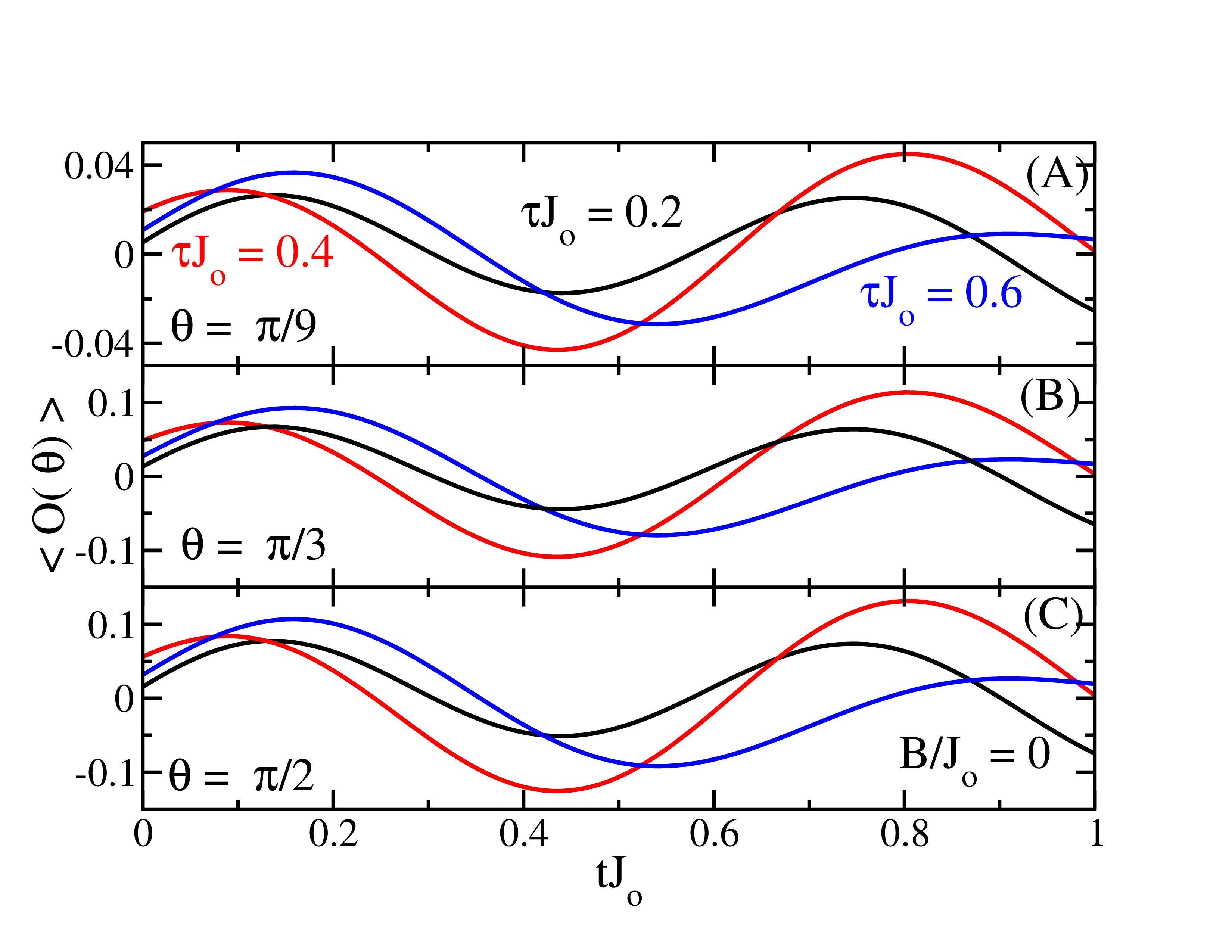}
	\end{center}
	\textbf{\refstepcounter{figure} \label{fig:isingsignal} Figure \arabic{figure}.} {(Color online.) Three different examples of $\hat{\mathcal{O}}(\theta)$ as a function of time when $\tau = 0.2$ (black), $0.4$ (red), and $0.6$ (blue) for $\theta = \pi/6$ (A), $\pi/3$ (B), and $\pi/2$ (C) at $B/J_o = 0$. }
\end{figure}
We next show simulated data for the transverse field Ising model with $J_{\pm} = 1$ and $J_0 = 1$kHz. The parameters for the $J_{ij}$ are $\mu = 1.0219 \omega_{COM}$ and the antisymmetric ratio of the trap frequencies  is $0.691/4.8$ which results in an $\alpha \approx 1.0$. The initial state is evolved to $t_{stop} = 6\tau$ and before the time interval $t_{meas.}$ the field is quenched to zero, as shown in Fig.~\ref{fig:Isingcartoon}(A).

 In Fig.~\ref{fig:isingsignal}, we show the time evolution of $\mathcal{O}(\theta)$ with $\theta = \pi/9$, $\pi/3$, and $\pi/2$ for $3$ different $\tau J_0 =0.2$, $0.4$, and $0.6$. The amplitude of the oscillations follow a similar trend to Fig.~\ref{fig:lzsignal}, such that at $\tau =0.4$ the amplitude of the oscillations are at a maximum in comparison to other $\tau$'s. Additionally, as $\theta$ is increased to $\pi/2$ the amplitude of the oscillations increase as previously seen in the Landau-Zener example.      

\begin{figure}[h!]
	\begin{center}
	\includegraphics[scale=0.075]{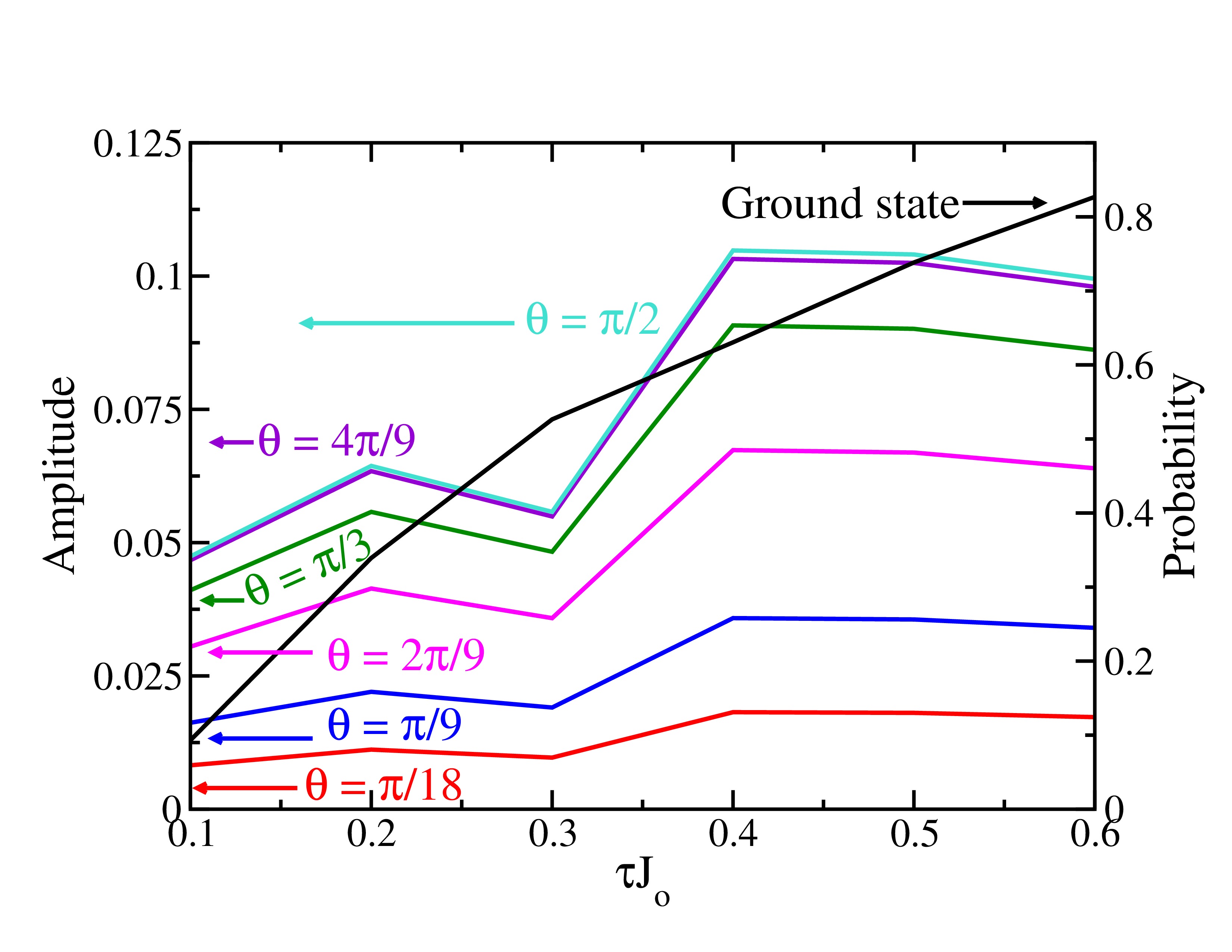}
	\end{center}
	\textbf{\refstepcounter{figure} \label{fig:isingamplitude} Figure \arabic{figure}.} {(Color online.)  Amplitude of the oscillations at $B/J_o = 0$ as a function of different $\tau$ as compared to the ground state probability. The arrows for each curve point toward the appropriate vertical axis. As the $\theta$ increases, the amplitude of the oscillations increases as well. The amplitude of the oscillations become a maximum when the ground state probability is near $0.6$ and the amplitude decreases when the ground state probability increases above or decreases below $0.6$. When $\tau = 0.3$ there is a local minimum in the amplitude. }
\end{figure}

In Fig.~\ref{fig:isingamplitude}, we compare the probability of the ground state to the amplitude as a function of the ramping $\tau$. In general, the amplitude of the oscillations is maximized near $\tau = 0.4$ when the ground state probability is $\approx 0.61$ and the amplitude decreases as the probability to be in the ground state either increases or decreases. Similar to the Landau-Zener problem, as the probability to be in the ground state increases to $1$ the amplitude will decrease to $0$. However when the ground state probability approaches $0.5$, the amplitude of the oscillations is at a local minimum. As previously seen in the Landau-Zener example, a single measurement of the amplitude ramped at $\tau$ cannot determine whether the probability of the ground state is high or low.       Hence, the probability needs to be tracked by using a series of measurements.

\begin{figure}[h!]
	\begin{center}
	\includegraphics[scale=0.075]{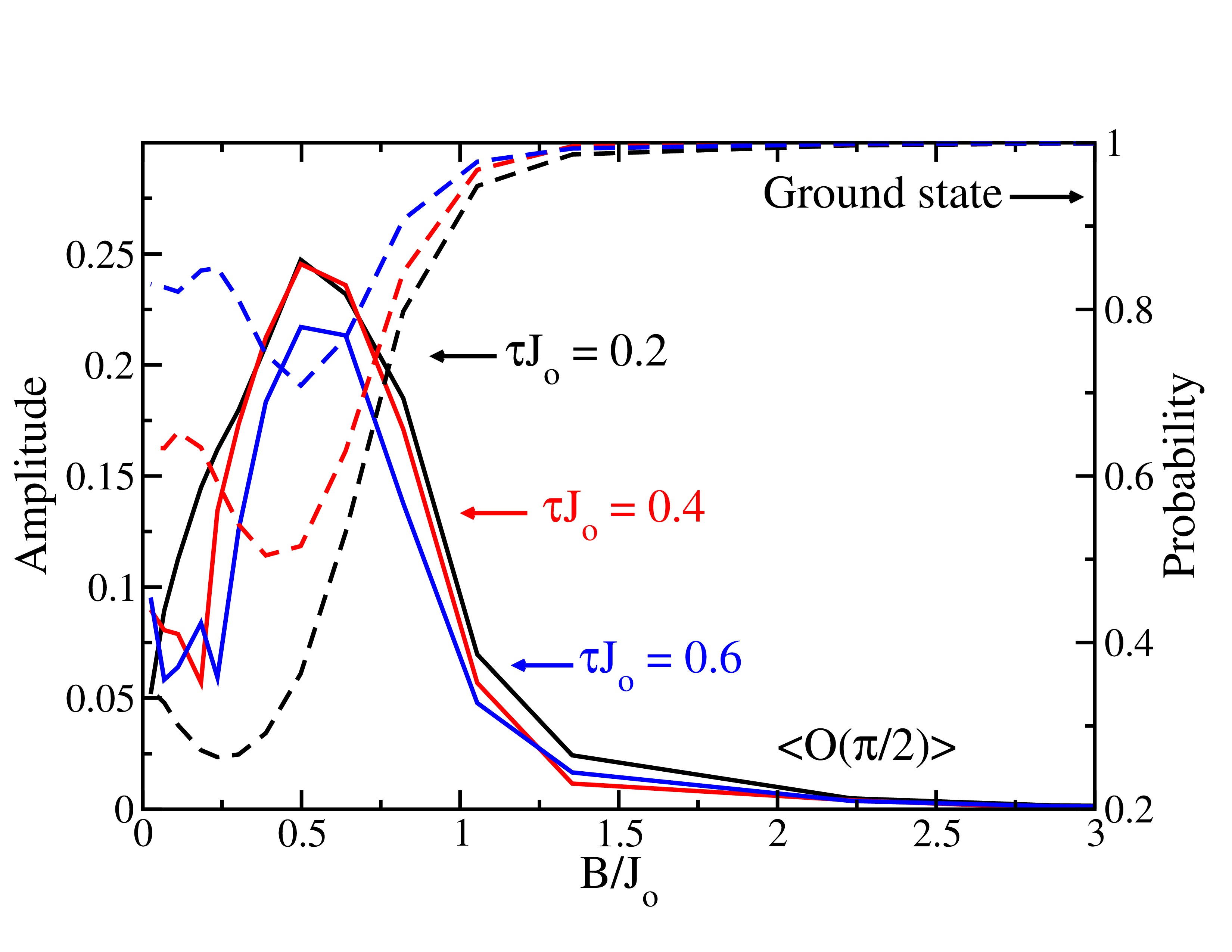}
	\end{center}
 \textbf{\refstepcounter{figure} \label{fig:isingrunning} Figure \arabic{figure}.} {(Color online.) Amplitude of the oscillations when the transverse field Ising model is held at different $B/J_o$ for $3$ different $\tau = 0.2$ (black), $0.4$ (red), and $0.6$ (blue), (where the arrows point toward the appropriate vertical axis). Before the minimum energy gap, the oscillations increase as the probability of the ground state decreases. However, after the minimum energy gap is passed,  depletion of the excited states back to the ground state occurs and the amplitude of the oscillations decrease accordingly. This depletion is difficult to detect by measuring the amplitude of the oscillations at only one time.}
\end{figure}

In general, the analysis of the amplitude can be done for different $t_{stop}$'s in which the transverse magnetic field is held constant at the strength of $B^x(t_{stop})$, shown in Fig.~\ref{fig:Isingcartoon}(B). Fig.~\ref{fig:isingrunning} shows the amplitude of the oscillations as a function of $B^x(t_{stop})$ for $3$ different $\tau = 0.2$, $0.4$, and $0.6$. The amplitude of the oscillations increases as the transverse magnetic field approaches the minimum energy gap and the excitations are created from the ground state, depending on the $\tau$. Once past the minimum energy gap, the ground-state probability increases as de-excitations occur and conversely the amplitude of the oscillations decrease as well. However a similar response will occur if excitations are being created after the minimum energy gap. Unfortunately, the analysis of the amplitude will not distinguish between these two possibilities. 

\section{Conclusion} 

In this work, we have proposed to analyze the amplitude of the oscillations for a given time-dependent Hamiltonian that is held constant for a time interval $t_{meas.}$ to extract information about the ground-state probability. We demonstrated this analysis for the Landau-Zener problem and for the transverse field Ising model (as would be simulated in the linear Paul trap). In both of the Hamiltonians, the amplitude of the oscillations becomes a maximum at a particular probability of the ground state and decreases as the ground state probability either increases or decreases. Hence a single measurement of the amplitude cannot determine which side of the maximum one is on. Therefore multiple measurements are needed to be made where the amount of excitations are varied. Additionally as the probability of the ground state is approaching $1$, the amplitude decreases to $0$ which can be difficult to measure given experimental noise. 

In this work, we have described the simplest analysis one can do to extract information about the probability of the ground state. This approach can be refined by using signal processing techniques like compressive sensing~\cite{donoh2006} to determine the Fourier spectra of the excitations. By monitoring the change of the weights of the delta functions, one can produce more accurate quantitative predictions for the probability of the ground state, because we can directly measure $P_1^*P_m$ for a few different $m$ values. But this goes beyond the analysis we have done here.

For the transverse field Ising model, de-excitations are observed, and were reflected in the amplitude of oscillations. However, after the minimum energy gap, more diabatic excitation can be created, but it is difficult to distinguish between the de-excitations and excitations.  One interesting aspect is that as long as the ground-state probability remains high enough, measuring the height of the oscillation amplitude can be used to optimize the ground-state probability as a function of parameters used to determine the time-evolution of the system. This can be a valuable tool for optimizing the adiabatic state preparation protocol over some set of opimization parameters.

\section*{Author Contributions}

JF and BY contributed equally to this manuscript.

\section*{Acknowledgments}

J. K. F. and B. T. Y. acknowledge support from the National Science Foundation under grant number PHY-1314295. J. K. F. also acknowledges support from the McDevitt bequest at Georgetown University. B.T.  Y. acknowledges support from the Achievement Rewards for College Students Foundation.

\bibliographystyle{frontiersinHLTH&FPHY} 
\bibliography{frontier}





\end{document}